# Room-temperature, continuous wave lasing in planar microcavities with quantum dots

Andrey Babichev,[1,*] Mikhail Bobrov,[1] Alexey Vasil'ev,[1] Sergey Blokhin,[1] Nikolay Maleev,[1] Ivan Makhov,[2] Natalia Kryzhanovskaya,[2] Leonid Karachinsky,[3] Innokenty Novikov,[3] and Anton Egorov[3]
[1]Ioffe Institute, Saint Petersburg, Russia
[2]HSE University, Saint Petersburg, Russia
[3]ITMO University, Saint Petersburg, Russia
*a.babichev@mail.ioffe.ru

Abstract
High-quality planar cavities with low-absorption mirrors based on $Al_{0.2}Ga_{0.8}As/Al_{0.9}Ga_{0.1}As$ layers demonstrate continuous wave lasing at a wavelength of 956 nm. At 300 K, the threshold power density and quality-factor at the threshold are (4.2±0.3) kW/cm$^2$ and (6800±220). Increasing the pump level above two thresholds lead to an enlargement in the quality-factor to at least 19000. Efficient lateral heat dissipation in the planar semiconductor microcavity is confirmed by a low mode-energy shift of approximately 400 μeV at two lasing thresholds.

Active regions based on quantum dots (QDs) have demonstrated their effectiveness in the fabrication of microdisk[1], dual-state lasers[2], single-photon sources[3], and vertical microcavity lasers[4-9].

High-speed spin-lasers[10] based on bimodal QDs micropillar cavities have recently been demonstrated. Compared with birefringent vertical-cavity surface-emitting lasers (VCSELs)[10] and optically pumped microlasers[11], micropillar cavity lasers with a large quality-factor (Q-factor) are of interest for quantum nanophotonics[12] and neuromorphic computing (NC)[13]. Reservoir computing (RC) is one of the NC schemes. The RC concept, based on diffraction-coupled laser arrays, has been presented for both VCSELs[14–17] and optically pumped microlasers[12]. An optical reservoir computer based on current-injected VCSELs (24 physical nodes) operating at room-temperature has been implemented[14].

The prospects for using optically pumped micropillar cavities as a RC elements are determined by their small substrate pitch[12] and the ability to operate at room temperature[18]. Achieving a large Q-factor requires an optimized vertical deep (~7–10 μm) dry etching process.

To mitigate deep etching of micropillar cavities, a concept of vertical hybrid photonic-defect microcavities with embedded QDs was invented[8]. Due to the quasi-planar geometry, this concept mitigates non-radiative surface recombination on the sidewalls, which is crucial when using active regions based on quantum wells (QWs)[12]. This cavity concept can be applied to fabricate the ultra-dense laser array required for RC.

Photonic-defect microcavities[7,8] consist of a semiconductor bottom mirror, a GaAs cavity with embedded QDs, and a dielectric output mirror deposited on the partially etched mesa. A micro-curved dielectric mirror, which was also used in the fabrication of blue GaN-based VCSELs[19], functions as a photonic-defect[7,8].

To mitigate the second epitaxy process, a dielectric top mirror was used[20]. Compared with the mesa epitaxial regrowth (buried mesa), which has been shown to be effective in wafer-fusion[21–25] it can be easily deposited/sputtered. The large refractive index contrast of dielectrics allows to reduce the number of mirror pairs and expand the reflectivity stopband from 70 (semiconductor mirrors) to 190 nm. The bottom mirror included GaAs/Al0.9Ga0.1As layers, which are characterized by a larger refractive index contrast[20], but have significantly greater loss at the pump wavelength compared to $Al_{0.2}Ga_{0.8}As/Al_{0.9}Ga_{0.1}As$ mirrors. Compared to micropillar cavities, the active region layers are not etched, and photonic-defect microcavities based on QWs are also possible[26]. Furthermore, the quasi-planar microcavities concept provides efficient lateral heat dissipation[7,8].

The discussed advantages of photonic-defects cavities (absence of non-radiative surface recombination and efficient lateral heat dissipation) are also characteristic of planar cavities. Q-factor simulations were performed for these two cavity types[7,8]. The transition from planar to quasi-planar (photonic-defect) cavities can positively affect the Q-factor when using a design with a shallow mesa[8]. On the contrary, the use of a deep mesa leads to the opposite effect (a decrease in Q-factor compared to a planar cavity[8]).

In the first experiment with quasi-planar cavities[8], the etch depth was 100 nm, which resulted in a modest increase in the Q-factor from 10500 (planar cavity) to 11000 (photonic- defect cavity). The maximum operating temperature was 260 K, and the threshold power density was 512 kW/cm$^2$ (16.1 mW)[8]. Increasing the etching depth to 120 nm, as well as the photonic-defect diameter, made it possible to operate at room temperature[7]. The threshold power density for a cavity with a photonic-defect diameter of 5 μm was 156 kW/cm$^2$ (4.9 mW)[7].

Quasi-planar cavities based on InGaAs QWs have also been implemented[26]. The bottom and top mirrors consist of 30 pairs of AlAs/GaAs and 11 pairs of $SiO_2/HfO_2$. The threshold power density for a cavity with a photonic-defect diameter of 0.75 and 2 μm was 25.5 and 2.5 kW/cm$^2$ (5.0 and 0.5 mW)[26].

In comparison to photonic-defect cavities, for which the index guiding is determined by the mesa profile, the mode-guiding in planar cavities is mainly determined by the thermal lens effect.

Compared to micropillar cavities, planar cavities mitigate sidewall loss, resulting in larger Q-factor. Furthermore, this concept does not require deep mesa etching. Compared to photonic-defect cavities, planar cavities do not require mesa

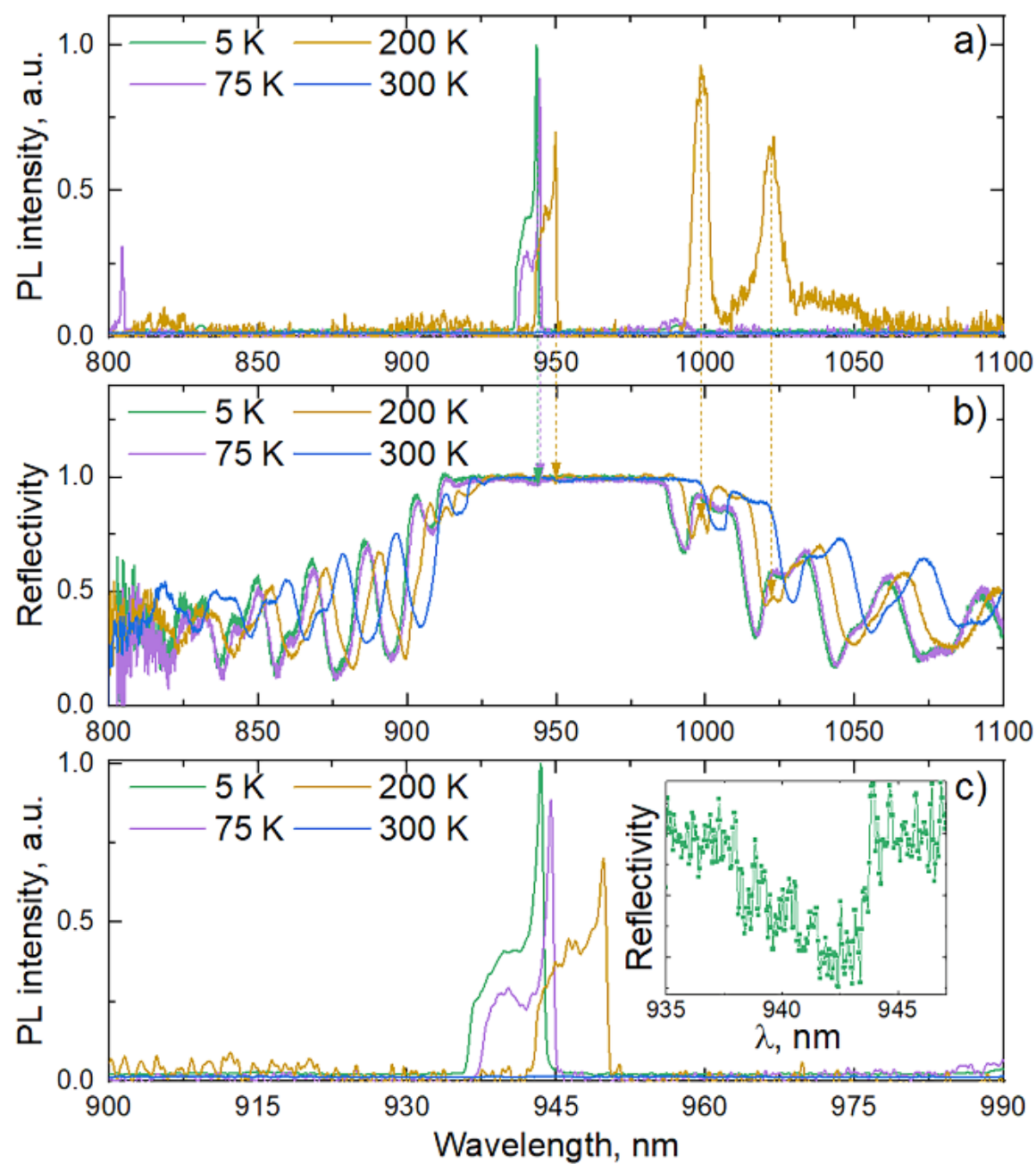

Figure 1. (a) and (c) PL spectra (normalized) of a planar cavity structure with QDs, measured from the surface of the structure; (b) Reflection spectra of a planar cavity structure with QDs, measured from the surface of the structure. The inset in panel (c) shows an enlarged reflection spectrum measured at 5 K.

etching, mitigating the scattering and defect-state absorption effects[7].

Planar GaAs/AlAs cavities with QWs[27] and quantum boxes (QBs)[28] have been examined previously. Due to the use of highly absorbing (at a wavelength of 514 nm) GaAs/AlAs mirrors, lasing in a planar vertical microcavity based on an $In_{0.14}Ga_{0.86}As$ QW was not observed[27]. The use of an opposite pumping scheme (wetting layers, WLs, pumping, at a wavelength of 826 nm), mitigates absorption inside the GaAs/AlAs mirror layers. At the same time, due to the small WLs thickness, the power conversion efficiency (PCE) is less than one percent[29], which limited lasing in planar GaAs/AlAs cavities with InAs QBs[28].

In this Letter, we present the first results of a study of planar vertical microcavities based on $Al_{0.2}Ga_{0.8}As/Al_{0.9}Ga_{0.1}As$ mirrors operating at room temperature, which is important for the practical application of microcavity lasers.

Molecular-beam epitaxy (MBE) was used to grow the planar cavity structure. A one lambda GaAs cavity was formed using low-absorption (at the pump wavelength) mirrors based on quarter-wave $Al_{0.2}Ga_{0.8}As/Al_{0.9}Ga_{0.1}As$ layers[30]. To fabricate a high-quality semiconductor microcavity structure, the bottom and top mirrors include 37.5 and 32 pairs. The modelled Q-factor of the cold cavity (excluding active region absorption), determined at 300 K, is 82400. The Stranski-Krastanow growth method was used to self-assemble the initial InGaAs QDs[18]. To increase the modal gain of the active region, three stacked layers of QDs were used. Barriers made of 20 nm thick GaAs were used. Optimizing epitaxy conditions made it possible to mitigate vertical alignment[18,30]. The gain to cavity detuning is zero at 5 K and increases to 72 nm at 300 K.30

For optical characterization, the planar cavity structure was measured in a Cryostation® s50 closed-cycle optical cryostat, which provides exceptional thermal and vibrational stability. Optical pumping was performed in continuous wave mode. A Mitutoyo MPlan Apo NIR 20× microscope was used to focus the emission on the sample. Surface emission spectra were collected using the same microobjective used for pumping. An Andor Shamrock 500i spectrometer (with a focal length of 500 mm) was applied to measure the emission spectra. A back- illuminated silicon CCD detector (Andor iDus 401A) was used to signal detection. The spectrometer was coupled with a 1200 lines/mm diffraction grating, providing a resolution of 0.05 nm. Reflection/spontaneous emission spectra from the planar cavity structure (see Fig. 1) were excited and collected using a microobjective similar to that used in the stimulated emission spectra experiment (see Fig. 2). The objective was a Mitutoyo MPlan Apo NIR 50×. The pump power was 95 μW at 5 and 75 K, increasing to 700 μW at 200 and 300 K.

First, we present the results of photoluminescence (PL) spectra measurements for a planar cavity structure pumped at a wavelength of 527 nm (see Fig. 1(a) and (c)). At 5 K, the PL spectrum exhibits a high-intensity peak near 943 nm, the position of which coincides with the position of the dip in the reflectivity stopband (see Fig. 1(b)). An additional low-intensity peak is observed near 990 nm (see Fig. 1(a)), which is associated with the emission through the fringe in the reflection spectrum.

Increasing the temperature to 75 K shifts of the main PL peak to approximately 945 nm (see Fig. 1(c)). The main peak also exhibits an additional short-wavelength shoulder (in the range of 937–945 nm), similar to the results at 5 K. This shoulder is associated with absorption in the active region, which also leads to an asymmetric dip in the reflection spectrum (see Fig. 1(c), inset). The emission through the fringe in the reflection spectrum corresponds to a wavelength of 990 nm.

At 200 K, the main PL peak (at the resonant wavelength of the cavity) is located at a wavelength of 950 nm (see Fig. 1(c)). Compared with the results obtained at low temperatures (at 5 and 75 K), two emission peaks are observed, corresponding to the emission through fringes in the reflection spectrum (at 998 and 1022 nm, see Fig. 1(b)). The reason for this is an increase in the pump power density from 3.0 kW/cm$^2$ (spectra measured at 5 and 75 K) to 22.3 kW/cm$^2$. The intensity of the peaks corresponding to the interference fringes in the reflection spectrum exceeds the intensity of the main PL peak (at the resonant wavelength of the cavity). This is explained by the smaller temperature shift of the resonant wavelength compared to the temperature shift of the peak from the QDs. A further increase in temperature to 300 K results in the absence of a spontaneous emission signal detected through the output mirror at a given pump power density of 22.3 kW/cm$^2$.

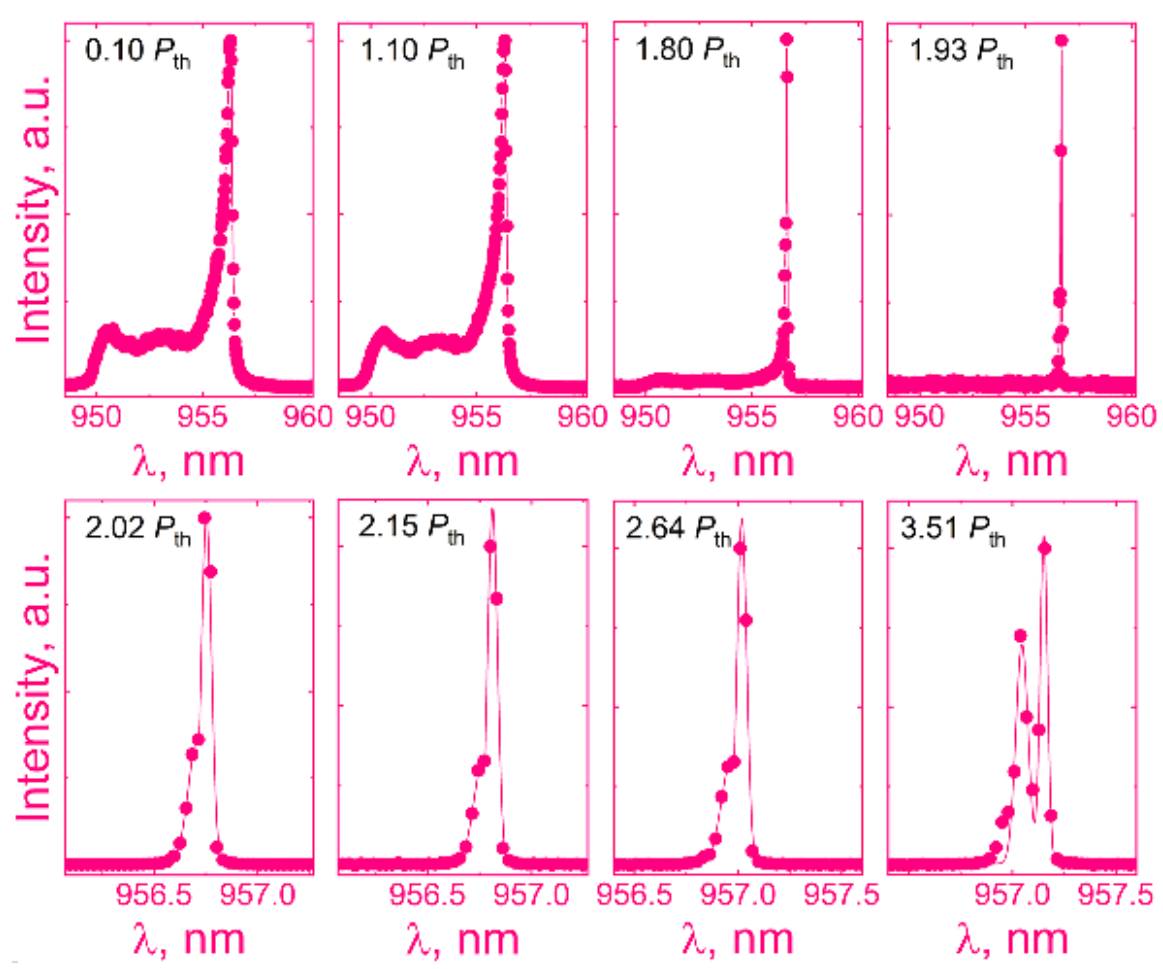

Figure 2. Normalized emission spectra of a planar microcavity structure measured at 300 K. The bottom panels (in the pump power density range 2.02–3.51×$P_{th}$) correspond to an enlarged scale along the X-axis.

To clarify the reason of the absence of emission at a pump power density twice the threshold level determined at a pump wavelength of 808 nm (see below), it is necessary to estimate the PCE value at both wavelengths. This value is determined by the transmission of the output mirror (Tmirror(λ)) and the absorption of the active region (A(λ))[31]. The absorption coefficient of $Al_{0.2}Ga_{0.8}As$ at 527 nm is 63604 cm$^{-1}$, which is approximately 570 times larger than that at 808 nm[32]. Absorption within the $Al_{0.9}Ga_{0.1}As$ layers is observed only at the pump wavelengths of 527 nm (~900 cm$^{-1}$)[32]. As a result, optical loss in the mirrors are mainly associated with the $Al_{0.2}Ga_{0.8}As$ layers.

Taking into account the absorption coefficients of the mirror layers ($\alpha_{AlGaAs}(\lambda,T)$), the value of Tmirror can be estimated using the expression[29]: $T_{mirror} = (1 - R(\lambda))\exp(-\alpha_{AlGaAs}(\lambda,T)d_{mirror})$, where R(λ) is the reflectivity at the wavelength λ and $d_{mirror}$ is the total thickness of the mirror layers. The value of A(λ) is determined according to $A(\lambda)=1 - \exp(-\alpha_{GaAs}(\lambda,T)d_{AR})$, where $\alpha_{GaAs}$ is the absorption coefficient of the cavity layer (GaAs) and $d_{AR}$ is the cavity thickness[29]. As a result, the use of pumping at short wavelengths leads to an increase in the absorption coefficient of GaAs and, consequently, to an enlarge in the absorption of the active region. The latter value can positively impact power conversion efficiency. Since the αGaAs value determined at 527 nm is 82844 cm$^{-1}$[32], which is 7.5 times larger than that at 808 nm, the extracted A(λ) value increases to 0.82 compared to 0.21 at 808 nm. Despite this positive effect, the contribution of the output mirror transmittance remains dominant in the case of strong absorption in its layers. The PCE value extracted at 527 nm is 3.75×10$^{-5}$ %, compared to 14.9 % at 808 nm. This is the main factor limiting laser action in planar cavities pumped at 527 nm.

Figure 2 shows the emission spectra measured at 300 K, which corresponds to a pump wavelength of 808 nm. Against the background of spontaneous emission (in the range of 949–957 nm), a superlinear increase in the mode intensity (at ~956 nm) is observed with a raise in the pump power.

It has been previously shown that the pronounced *S*-shape of the input-output (I-O) curve, along with linewidth narrowing, indicates an underlying laser transition for micropillar[31] and photonic-defect[7,8] microcavities. Photon autocorrelation studies can be used to provide further evidence for the laser transition[7].

The luminescence line (~956 nm) at different pump levels was approximated (Gauss line shape fitting[29]). The emission intensity was analyzed by solving the laser rate-equations[11]. The dependence of the pump power density P on the number of photons in the microcavity n is described by the expression[11]: $P(n)=A[n(1+\xi)(1+\beta n)/(1+n)-\xi\beta n]$, where ξ (8×10$^{-3}$) is a scaling factor that takes into account the number of excitons at transparency, β is the spontaneous emission factor and *A* is the scaling factor. The result of fitting the I-O characteristic defined using this expression, is shown in Fig. 3. The estimated β-factor is (2.2±0.13)×10$^{-3}$%. This value is 4 times smaller than that of the photonic-defect cavity (see Table 1). The modelled Q-factor for the examined cavity is more than two times smaller (see discussion below) than that of a planar cavity based on 33.5 pairs of $GaAs/Al_{0.9}Ga_{0.1}As$ and 19 pairs of $SiO_x/SiN_x$ mirrors[7]. This fact, along with a large mode volume compared to a photonic-defect cavity, leads to a smaller Purcell and β-factors. The threshold pump density $P_{th}$ is derived from the reduced expression: $P(n)=A(1+2\xi+2\beta\times(1-\xi))/3\beta$ and is equal to (4.16±0.26) kW/cm$^2$. The threshold absorbed power density, determined by multiplying the PCE value by the threshold power density, is (620±40) W/cm$^2$. The Q-factor at the threshold is 6800±220. Increasing the pump power density to 2.0×$P_{th}$ leads to a decrease in the linewidth (increase in Q-factor) to 0.05 nm (19000), which is limited by the spectral resolution of the monochromator used. The discrepancy between the measured and modelled Q-factor is due to the absorption in the active region. A further increase in the pump power density leads to a gain in the intensity of the additional mode (see Fig. 2), which causes a decrease in the intensity of the fundamental mode and an increase

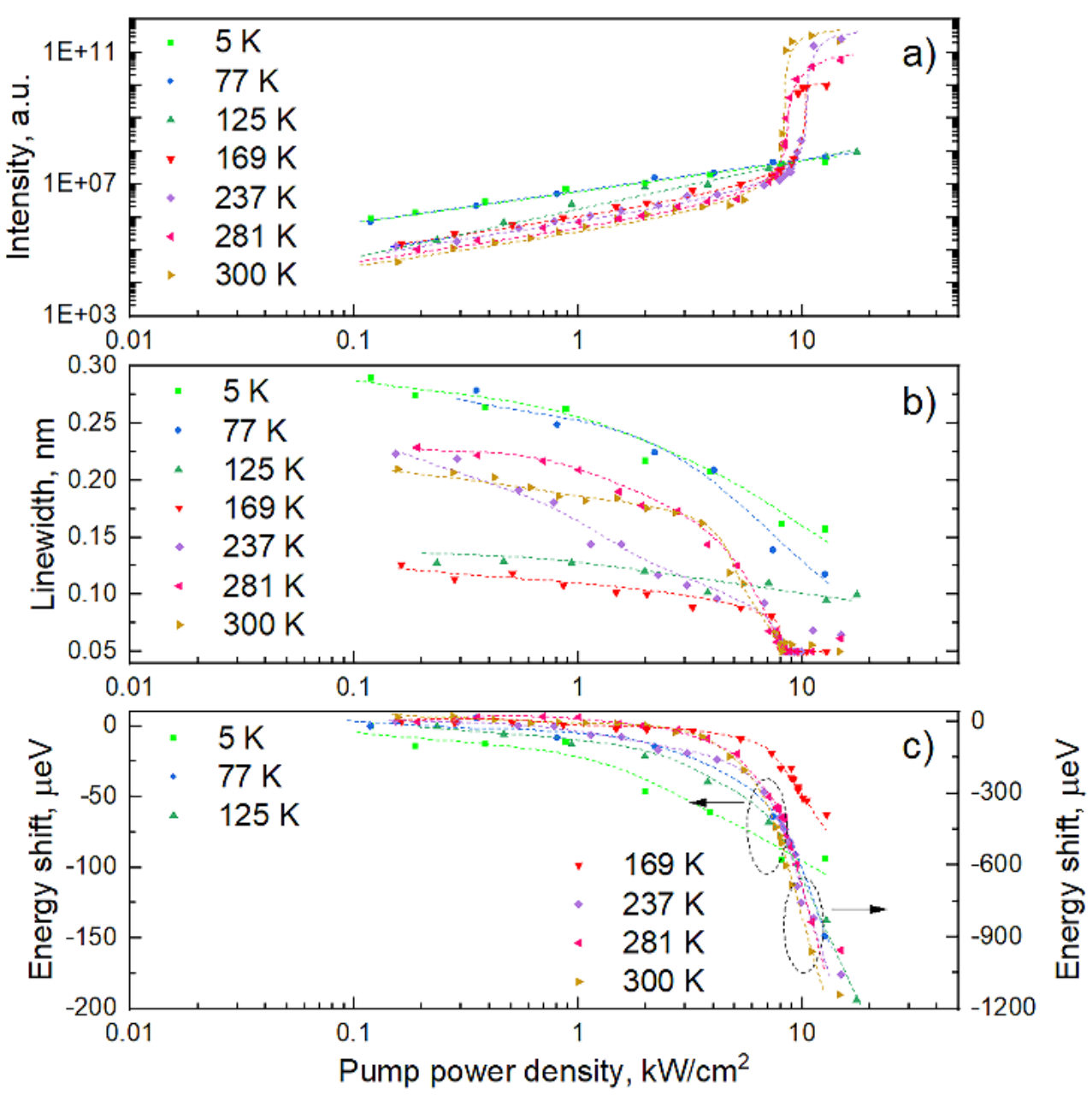

Figure 3. (a) I-O characteristics of the planar microcavity structure determined in the temperature range 5–300 K; (b) Linewidth at different pump power densities determined in the temperature range 5–300 K; (c) Mode energy shift as a function of pump power, determined in the temperature range 5–300 K. The left Y-axis corresponds to temperatures of 5, 77 and 125 K. The right Y- axis corresponds to temperatures of 169, 237, 281 and 300 K.

in its linewidth (see Fig. 3).

An increase in the mode spacing with increasing pump level is a general characteristic of microcavities with thermally induced index guiding[33]. This effect is demonstrated for the examined planar cavities; increasing the pump power density from 8.0 to 14.6 kW/cm$^2$ leads to an enlarge in the mode spacing from 57×10$^{-3}$ to 102×10$^{-3}$ nm (see Fig. 2).

To analyze thermal issues, the mode energy shift (ΔE) is commonly analyzed[31], which is shown in Fig. 3(c). The redshift of the mode energy is demonstrated over the entire pumping range, indicating the predominance of the thermal over the plasma effect[31]. This fact allows us to conclude that the thermal lens effect[34] is the dominant compared to the gain guiding caused by the increase in the charge carrier density[33]. Increasing the pump power density from 0.1(1.0)×$P_{th}$ to 2.0×$P_{th}$ results in a redshift of ΔE value by approximately 580 (410) μeV. Raising the pump power density from 0.1(1.0) to 3.5×$P_{th}$ leads to a mode energy shift of approximately 1.14 meV (1.04 meV).

For comparative analysis, the mode energy shift of micropillar lasers operating at room temperature[18] should be discussed (see Table 1). The lasing threshold for a 5 μm-diameter pillar corresponds to a threshold power density of 6.2 (semiconductor output mirror) and 8.8 (hybrid output mirror) kW/cm$^2$[18]. With an increase in the pump power density from 1.0 to 2.0×$P_{th}$, a mode energy shift of –2.3 (semiconductor output mirror) and –1.4 (hybrid output mirror) meV was extracted[18].

Compared with hybrid photonic-defect cavities using a non-absorbing dielectric output mirror,7 the use of a low-absorption output mirror (based on $Al_{0.2}Ga_{0.8}As/Al_{0.9}Ga_{0.1}As$ layers) leads to additional heating, which has also been observed in micropillar cavity lasers[7, 18]. As a result, the thermal lens effect[34] (the dominant effect over the gain guiding) makes it possible to implement optical confinement in planar cavities based on $Al_{0.2}Ga_{0.8}As/Al_{0.9}Ga_{0.1}As$ mirrors. In the examined planar cavity, increasing the pump power from 0.1 to 1.0×$P_{th}$ leads to a mode energy shift of 100 μeV (0.078 nm). Based on the temperature shift of the mode position (0.055 nm/K), the heating with increasing pump power from 0.1 to 1.0(2.0)× $P_{th}$ is approximately 1.4(7.8)°C. Taking into account the temperature shift of the refractive index (4×10$^{-4}$ K$^{-1}$)[34], increasing the pump level from 0.1 to 1.0(2.0)× $P_{th}$ provides a lateral refractive index ($\Delta n_{lat}$) contrast of 0.6(3.8)×10$^{-3}$. A further rise in the pump power density to 3.5×$P_{th}$ leads to a $\Delta n_{lat}$ value of 5.4×10$^{-3}$.

Temperature studies were conducted to further confirm that the thermal lens effect is the primary mechanism responsible for lasing in the examined planar microcavities (see Fig. 3). It is predicted that lowering the temperature will reduce the heating and, consequently, the lateral confinement caused by the thermal lens effect.

At 5, 77 and 125 K, a linear I-O characteristic was demonstrated. At 5 K, increasing the pump power density from 0.12 to 12.6 kW/cm$^2$ results in a redshift of the mode energy of only 94 μeV. The linewidth is 0.290 nm and decreases to 0.157 nm with increasing pump level. At temperatures of 77 K and 125 K, increasing the pump level from 0.12 to 12.6 kW/cm$^2$ leads to a redshift of the mode energy by 149 and 222 μeV, respectively. The minimum linewidth (at 12.6 kW/cm$^2$) extracted at 77 K and 125 K is 0.117 and 0.100 nm.

Increasing the temperature to 169 K results in an *S*-shaped I-O characteristic (double logarithmic scale) and a narrowing of the linewidth (to 0.05 nm at 8.0 kW/cm$^2$ and above, see Fig. 3(b)). These facts confirm the presence of lasing. The estimated lasing threshold is (5.38±0.32) kW/cm$^2$. At this pump level, the redshift of the mode energy is 79 μeV and increases to 390 μeV at a pump power density of 12.6 kW/cm$^2$. These ΔE values coincide with the estimated heating of 1 (at 5.4 kW/cm$^2$) and 5 ºC (at 12.6 kW/cm$^2$).

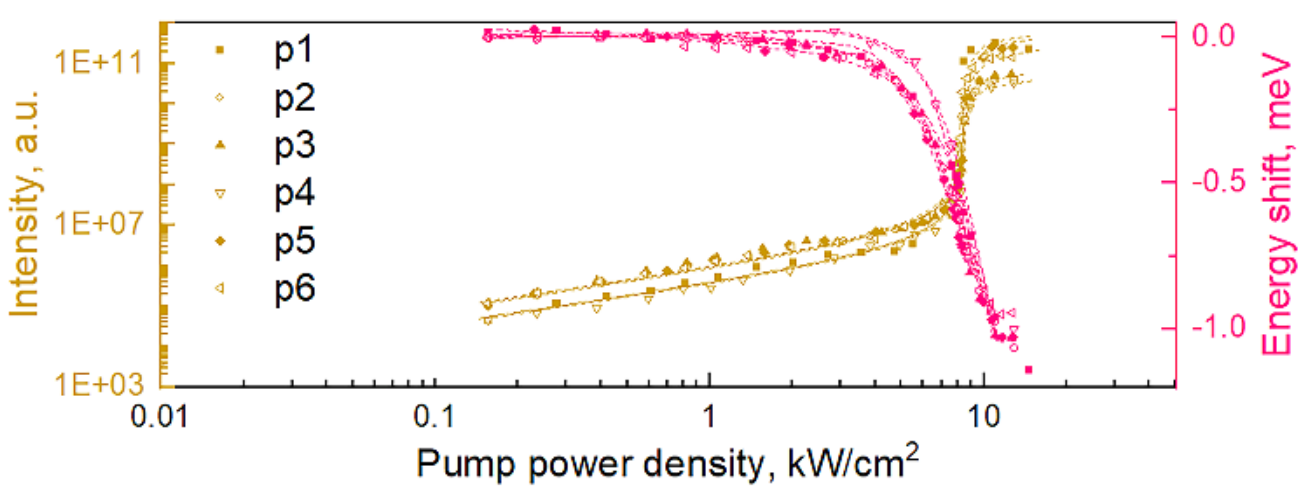

Figure 4. I-O characteristics of the planar microcavity structure measured at different points (left Y-axis); Mode energy shift as a function of pump power measured at different points (right Y-axis).

At 237 K, a decrease in the lasing threshold to increase in the lateral optical confinement caused by large (5.24±0.31) kW/cm2 was demonstrated, as a result of an thermal heating (thermal lens effect). The ΔE value is 920 μeV at a pump power density of 12.6 kW/cm2.

Increasing the temperature to 281 K leads to a decrease in the lasing threshold to (4.30±0.26) kW/cm2. Increasing the pump power density to two thresholds and above results in a Q-factor more of more than 19000. The mode energy shift is 990 μeV at a pump power density of 12.6 kW/cm2.

Although the lasing threshold is larger at low temperatures, the Q-factor at the lasing threshold is larger. This Q-factor at temperatures of 169, 237, 281 and 300 K is (10900±340), (10200±320), (7020±220) and (6800±220). The similar behavior has been previously discussed for micropillar cavities[30] and may be associated with the lower absorption in the active region in the low temperature region.

The wavelength just above the threshold is determined. At 169 K, this value is 949.2 nm. Increasing the temperature to 300 K results in the lasing with a wavelength of 956.4 nm. In the temperature range of 169–300 K, a linear temperature shift of the wavelength (at the lasing threshold) of 0.055 nm/K is observed. This value is close to the previously discussed temperature shift of the dip position in the reflection spectrum (0.058 nm/K)[30], and was used in the heating estimation discussed above.

To study the reproducibility over the substrate, the output characteristics of 6 planar microcavities were investigated over a square area of 3×3 mm2 (see Fig. 4). The maximum threshold power density is (4.28±0.26) kW/cm2, which coincides with the wavelength of 959.016 nm. At this measurement point, the β-factor is (0.6±0.02)×10-3%. The minimum threshold power density is (4.06±0.25) kW/cm2 corresponding to the maximum β-factor of (3.6±0.13)×10-3%. The wavelength determined just above the threshold is 958.291 nm.

In conclusion, the fabrication of semiconductor photonic-defect cavities is limited due to the asymmetric in-plane regrowth profile[35, 36]. A new type of room-temperature microlasers based on high-quality planar semiconductor microcavities is presented. Coherent lasing at room temperature is confirmed by a clear *S*-shaped I-O characteristic and linewidth narrowing[7]. At 300 K, the threshold power density is approximately 1.5 times lower than that of micropillar cavities[18]. The Q-factor at the threshold decreases to 6800 compared to 8100 for micropillar cavities[18]. The relative mode energy shift with doubling the lasing threshold is 5.6 times smaller compared to micropillar lasers[18] (see Table 1). As a result, the use of planar cavities allows for a significant reduction in thermal issues due to possibility of lateral heat dissipation.

Laser action of hybrid planar cavities is limited by the absence of absorption in the dielectric mirror[7]. In comparison to hybrid quasi-planar (photonic-defect) cavities based on QDs[7], the examined planar cavities demonstrate a 37-fold reduction in threshold power density. A similar Q-factor at threshold is extracted (5300[7]). The excellent thermal performance of quasi-planar cavities was confirmed by the low Raman shift (0.344 cm$^{-1}$)[7].

Hybrid photonic-defect cavities based on QWs[26] exhibit superlinear I-O characteristics (on a linear scale) and linewidth narrowing. The lower lasing thresholds than those studied here (see Table 1) may be due to the larger material gain of the QWs compared to QDs.

Further optimization of the planar semiconductor cavity (minimization of thermal issues) should be accompanied by an increase in the Q-factor. In fact, a twofold increase in the Q-factor for a micropillar cavity with a hybrid output mirror compared to the semiconductor one[18], results in a 1.6-fold decrease in the mode energy shift (see Table 1).

Similar behavior is observed for hybrid photonic-defect cavities, for which a redshift of the mode energy is observed in microcavities with a lower Q-factor (see Table 1). The estimated Q-factor for a planar hybrid cavity based on 33.5 pairs of $GaAs/Al_{0.9}Ga_{0.1}As$ and 19 pairs of $SiO_x/SiN_x$[7] is 25 times larger than for a cavity based on 30 pairs of GaAs/AlAs and 11 pairs of $SiO_2/HfO_2$[26].

Sputtering two pairs of $SiO_2/Ta_2O_5$ onto the examined planar cavity, similar to microlillar cavities[18], makes it possible to obtain a Q-factor close (0.76) to the value for a planar design based on 33.5 pairs of $GaAs/Al_{0.9}Ga_{0.1}As$ and 19 pairs of $SiO_x/SiN_x$ mirrors[7]. The efficiency (PCE value at 808 nm) for these two designs is estimated to be 19.5 %[18] and 18.8 %[7]. Although the design with a hybrid output mirror uses low absorption mirrors[18], the effect of absorption on efficiency is negligible (0.3%).

A brief outlook of the planar cavity concept based on low-absorption mirrors is presented. High-β lasing action has been demonstrated in photonic-defect cavities based on transition metal dichalcogenides, TMDCs)[37]. However, lasing action in planar cavities has not been demonstrated due to the low Q-factor (230) of the fabricated dielectric microcavity [37].

Graphene and TMDCs can be fabricated over large areas using chemical vapor deposition[38]. This together with the possibility of MBE on graphene[39], demonstrates the potential for fabrication high-quality semiconductor planar cavities based on TMDCs, other 2D-materials, or colloidal QDs.

The epitaxy conditions of buried stressor site-controlled QDs (SCQDs)[40] were also developed, which leads to low-temperature lasing in micropillar cavities based on $Al_{0.9}Ga_{0.1}As$/GaAs mirrors[41]. The lasing at room-temperature (RT) in the examined planar cavity based on low-absorption mirrors demonstrates the promise of quasi-planar photonic-defect cavities based on SCQDs[9] operating at RT. Electrically injected planar cavities based on SCQDs[42] may also be of interest.